\documentclass[reqno,12pt]{amsart}

\usepackage{enumerate}
\usepackage[margin=1in]{geometry}
\usepackage{enumitem}
\usepackage{ifpdf}
\usepackage{amsmath}
\usepackage{amsfonts}
\usepackage{amssymb}
\usepackage{amsaddr}
\usepackage{amsthm}
\usepackage{mathrsfs}
\usepackage{amsrefs}
\usepackage{amsfonts}
\usepackage{amssymb}
\usepackage{cite}
\usepackage[hyperfootnotes=false]{hyperref}
\usepackage[titletoc,title]{appendix}
\usepackage[dotinlabels]{titletoc}
\usepackage[all,cmtip]{xy}
\usepackage{nicefrac}
\usepackage{float}
\usepackage{verbatim}
\usepackage[multiple]{footmisc}
\usepackage{mathdots}
\usepackage{dcolumn}
\usepackage{tabu}
\usepackage[hang,small,bf]{caption}    

\usepackage{tikz}	
\usetikzlibrary{backgrounds,fit,decorations.pathreplacing}  

\newlength\Li \newlength\Lii 
\newcommand{\ignore}[1]{}

\newcommand{\abs}[1]{\left\lvert {#1} \right\rvert}
\newcommand{\norm}[1]{\left\lVert {#1} \right\rVert}

\newtheorem{thm}{Theorem}[section]

\newtheorem{cor}[thm]{Corollary}

\theoremstyle{definition}
\newtheorem{defn}[thm]{Definition}

\theoremstyle{remark}

\theoremstyle{note}

\usepackage{tikz}
\usetikzlibrary{backgrounds,fit,calc,trees,positioning,arrows,chains,shapes.geometric,%
    decorations.pathreplacing,decorations.pathmorphing,shapes,%
    matrix,shapes.symbols,snakes}

\newcommand{\ket}[1]{\ensuremath{\left|#1\right\rangle}} 
\newcommand{\bra}[1]{\ensuremath{\left\langle#1\right|}} 

\tikzstyle{sum} = [draw,circle,inner sep=0mm,minimum size=2mm]
\tikzstyle{connector} = [->,thick]
\tikzstyle{line} = [thick]
\tikzstyle{branch} = [circle,inner sep=0pt,minimum size=1mm,fill=black,draw=black]
\tikzstyle{guide} = []
\tikzstyle{snakeline} = [connector, decorate, decoration={pre length=0.2cm,
                         post length=0.2cm, snake, amplitude=.4mm,
                         segment length=2mm},thick, magenta, ->]

\makeatletter
\newcommand*{\rom}[1]{\expandafter\@slowromancap\romannumeral #1@}
\makeatother

\newtheoremstyle{defnopunct}
{3pt}
{3pt}
{}
{}
{\bfseries}
{ }
{.5em}
{}

\usepackage{etoolbox}
\let\bbordermatrix\bordermatrix
\patchcmd{\bbordermatrix}{8.75}{4.75}{}{}
\patchcmd{\bbordermatrix}{\left(}{\left[}{}{}
\patchcmd{\bbordermatrix}{\right)}{\right]}{}{}

\author{Ji\v{r}\'{\i} Lebl}
\thanks{Ji\v{r}\'{\i} Lebl was in part supported by Simons Foundation collaboration grant 710294.}
\address{Department of Mathematics,  Oklahoma State University, 
Stillwater,  OK 74078,  USA}
\email{lebl@okstate.edu}

\author{Asif Shakeel}
\address{CA 92122}
\email{asif.shakeel@gmail.com}
\ifpdf
\fi

\title[VQA for Euclidean Discrepancy and Covariate-Balancing] {Variational Quantum Algorithms for Euclidean Discrepancy and Covariate-Balancing}

\begin{document}

\begin{abstract}
Algorithmic discrepancy theory seeks efficient algorithms to find those two-colorings of a set that minimize a given measure of coloring imbalance in the set, its {\it discrepancy}.  The {\it Euclidean discrepancy} problem and the  problem of  balancing  covariates in randomized trials  have efficient randomized algorithms based on the  Gram-Schmidt walk (GSW).   We frame  these problems as  quantum Ising models, for which  variational quantum algorithms (VQA) are particularly useful.  Simulating an example of covariate-balancing  on an IBM quantum simulator, we find that the variational quantum eigensolver (VQE) and the quantum approximate optimization algorithm (QAOA) yield results comparable to the GSW algorithm.    

\end{abstract}


\maketitle

\section{Introduction} \label{sec:intro}
A broad description of discrepancy theory given in~\cite{jm:gdt}  is that it is the theory of the {\it irregularities of distributions}.  Precise formulations of discrepancy depend on  the spaces and sets being studied (for a comprehensive treatment, see~\cite{jm:gdt,bc:tdmrc}). To motivate this discussion, consider a finite discrete set $P$.  We are given a  collection $\mathcal{S}$  of subsets of $P$, constituting a set-system $(P,\mathcal{S})$. Let $\omega$ be a  {\it two-coloring} of $P$ given by a map from  $P$ to $\{+1,-1\}$,
\begin{equation*}
\omega: P \rightarrow \{+1,-1\}.
\end{equation*}
For a set $S \in \mathcal{S}$ and a coloring  $\omega$, let the  absolute  difference  in the number of elements in $S$ colored $+1$ and $-1$ be,
 \begin{equation*}
d_{S}(\omega) =  \abs{\sum_{p \in P \cap S} \omega(p)}.
\end{equation*}
Define the  {\it coloring-discrepancy} of the coloring $\omega$  as  the maximum of this difference, over all $S \in \mathcal{S}$, 
 \begin{equation*}
d_{\mathcal{S}}(\omega) = \max_{S \in \mathcal{S}} d_{S}(\omega).
\end{equation*}
Then the {\it discrepancy} of the set system  $(P,\mathcal{S})$  is defined as
\begin{equation*}
\text{disc} (\mathcal{S}) = \min_{\{\omega\}}  d_{\mathcal{S}}(\omega).
\end{equation*}
   Beck and Fiala~\cite{bf:imt}  established a highly regarded bound on discrepancy in discrete spaces.  Spencer~\cite{js:ssds} subsequently obtained  the famed ``six standard deviations" result. The definition of discrepancy is expanded from above in spaces with more structure.  B\'{a}r\'{a}ny and Grinberg~\cite{bg:oscqfds} derived results on vector spaces with discrepancy quantification via seminorms. Banaszczyk~\cite{wb:bftten} proved strong  bounds for Euclidean discrepancy and in~\cite{wb:bvgmndcb}  generalized the results of Beck and Fiala~\cite{bf:imt} to vector balancing in convex bodies. 
   
Recent accomplishments in algorithmic discrepancy theory realizing efficient algorithms for colorings which had previously been proven only to exist, began with Bansal's~\cite{nb:cadm} algorithm for Spencer's result~\cite{js:ssds}. It   allows efficient sampling of colorings with strongly bounded discrepancy, previously considered computationally hard. Bansal et al~\cite{bdgl:gswcbb} have also developed an efficient randomized algorithm, the Gram-Schmidt Walk (GSW) algorithm, that samples colorings in Banaszczyk's Theorem~\cite{wb:bvgmndcb} and several others. This has been refined to tighter bounds, and extended in both theoretical and algorithmic directions, in the work of Harshaw et al~\cite{hssz:bcregswd} on covariate balancing. 

In the current era of  Noisy Intermediate-Scale Quantum (NISQ) devices~\cite{p:qcnisq},
Variational Quantum  Algorithms (VQA), also called  hybrid
quantum-classical algorithms, find much use.  The main VQA are  the  Variational Quantum Eigensolvers (VQE)~\cite{pms:vqepqp,mrb:tvhqca}  and  the Quantum Approximate Optimization Algorithm (QAOA)~\cite{fgg:qaoa}, both of which  have been shown to be very effective   in approximating solutions to certain  optimization problems~\cite{yrs:ovqapmp}, and   approximating  the ground state energies (minimum eigenvalues) of Hamiltonians~\cite{crd:cmsqpea, nlc:oqaoasprdcd, pbb:qaoalrimtqs,zwc:qaoapmin}.

In the present paper, we focus on a restricted version of discrepancy in~\cite{bg:oscqfds, wb:bvgmndcb, hssz:bcregswd},  on a finite dimensional real vector space with Euclidean norm.  This is naturally suited to quantum algorithms which assume a complex vector space with inner product norm. It also lends itself to the connection with the Ising model~\cite{al:ifmnpp}, and thus to VQA.

The rest of this paper is organized as follows. Section~\ref{sec:eucdisc} introduces  Euclidean discrepancy and the theorem of Banaszczyk~\cite{wb:bftten} that sets bounds on it. Section~\ref{sec:vqaising} is a very brief description of the   variational quantum algorithms (VQA) and Hamiltonians for Ising models. Section~\ref{sec:quso} discusses the Quadratic Unconstrained Signed Optimization (QUSO).     Section~\ref{sec:isingeucdisc} sets up the Ising model for the Euclidean discrepancy and covariate-balancing from~\cite{hssz:bcregswd}, explaining the context of randomized trials. With this formulation, these problems are accessible by the VQA.    Section~\ref{sec:quantsim} then tests the widely used  variational quantum  algorithms,  Variational Quantum Eigensolvers (VQE)  and  the Quantum Approximate Optimization Algorithm (QAOA), on a quantum simulator to sample assignments for a test case. The experiments confirm the ability of these quantum algorithms, on that example, to sample  assignments that perform well relative to the GSW algorithm. Section~\ref{sec:conc} is the conclusion.

\section{Euclidean discrepancy} \label{sec:eucdisc}
In this setting, we have $\mathbb{R}^n$ as the ambient space, and a  set of $m$  vectors $X=\{x_1,\dots, x_m \in \mathbb{R}^n\}$. Let the coloring-discrepancy  be the map from the space of colorings $\{\omega_1,\ldots,\omega_m = \pm 1\}$ to $\mathbb{R}$,
 \begin{equation} \label{eucdiscfun}
d_X(\omega) = \norm{ \sum_{i=1}^m \omega_i x_i},
\end{equation}
where $\norm{\cdot}$ is the Euclidean norm. The Euclidean discrepancy of the set of vectors $X=\{x_1,\dots, x_m \in \mathbb{R}^n\}$ is  
 \begin{equation} \label{eucdisc}
\text{disc}(X) = \min_{\{\omega\}}d_X(\omega).
\end{equation}
A coloring $\omega^*$ is called an {\it optimal} coloring if 
 \begin{equation} \label{eucdiscopt}
d_X(\omega^*) = \text{disc}(X).
\end{equation}
 Let us recall a theorem of Banaszczyk~\cite{wb:bftten}. 
\begin{thm}[Banaszczyk~\cite{wb:bftten}] \label{Bthm0}
Let $D$ be an $n$-dimensional ellipsoid in $\mathbb{R}^n$ with center at $0$ and principal semi-axes $\lambda_1,\ldots, \lambda_n$. Choose any  $x_1,\dots, x_m \in D$ ($m$ is arbitrary, independent of $n$). Then:
 \begin{enumerate} [label=(\alph{*})] 
 \item \label{Bthm0a} there exist signs $\omega_1,\ldots,\omega_m = \pm 1$, such that
 \begin{equation*}
 \norm{ \sum_{i=1}^m \omega_i x_i} \le {\left( \sum_{j=1}^n   \lambda_j^2 \right)}^\frac{1}{2},
 \end{equation*}
  \item \label{Bthm0b}  to each $k = 1, ... ,m$ there corresponds a subset $I$ of $\{1, ... ,m\}$ consisting of exactly $k$ elements, such that
 \begin{equation*}
\norm{ \sum_{i \in I}x_i - \frac{k}{m}\sum_{i=1}^m  x_i} \le {\left( \sum_{j=1}^n   \lambda_j^2 \right)}^\frac{1}{2}.
 \end{equation*}
  \end{enumerate}
\end{thm}
In the above theorem, when the ellipsoid is a unit ball in $\mathbb{R}^n$,  the right side of the  inequalities become $\sqrt{n}$.~\footnote{Banaszczyk~\cite{wb:bftten} attributes this special case  to the earlier works of Sevastyanov~\cite{ss:oaspcp} and B\'{a}r\'{a}ny (unpublished).}


Let $X$ be the $n \times m$ matrix formed by using the vectors $\{x_i\}$ as column  vectors,
\begin{equation} \label{Xmat}
X=\begin{bmatrix} 
  \vline & \cdots & \vline  \\
     x_1& \cdots  & x_m\\
     \vline & \cdots  &  \vline 
  \end{bmatrix}.
\end{equation}
Let us also write $\omega_1, \ldots, \omega_m $ as a column vector $\omega$,
\begin{equation*}
\omega=\begin{bmatrix}
	\omega_1\\
    \vdots \\
     \omega_m
  \end{bmatrix}.
\end{equation*}
Then we can write  the square of coloring-discrepancy  in eq.~\eqref{eucdiscfun} as, 
\begin{equation} \label{eucdicfunmat}
d_X(\omega)^2 = \omega^{\top}X^{\top} X \omega.
\end{equation}

By the above theorem part~\ref{Bthm0a}, there exists a  vector $\omega \in \{-1,1\}^{\times m}$ such that $\norm{X \omega} \le \sqrt{n}$. This implies 
\begin{equation*} 
d_X(\omega)^2 =  \omega^{\top}X^{\top} X \omega \le n.
\end{equation*}

\section{Variational quantum algorithms and the Ising model} \label{sec:vqaising}
We give a short introduction to the Ising model and the VQA, the first being a core object in this paper, and the second being the paradigmatic tool that  realizes the optimization we need.  
  
Let $\mathcal{H}$  be a  Hilbert space of $m$ qubits,
\begin{equation} \label{hilbsp}
\mathcal{H}= {\mathbb{C}^2}^{\otimes m}.
\end{equation}
In the usual notation, let the standard basis vectors of $\mathbb{C}^2$ be
\begin{equation*}
\ket{0}=\begin{bmatrix}
    1\\
    0
  \end{bmatrix}, \quad
\ket{1}=\begin{bmatrix}
    0\\
    1
  \end{bmatrix}.
\end{equation*} 
 Then a natural  basis  of $\mathcal{H}$ is the  product-basis,
\begin{equation*} 
\mathcal{B}=\left\{{\bigotimes}_{i=1}^{m} \ket{b_i} : b_i =0,1\right\}.
\end{equation*}

A Hamiltonian $H$ is   a self-adjoint operator on $\mathcal{H}$, which, for the current discussion,  which is  expressible  as sum of  chains of Pauli spin matrices on $\mathcal{H}$.   By way of explanation, Pauli spin matrices are
\begin{equation*}
\sigma_x=\begin{bmatrix}
    0&1\\
    1 & 0
  \end{bmatrix}, \quad
  \sigma_y=\begin{bmatrix}
    0&-i\\
    i & 0
  \end{bmatrix}, \quad
  \sigma_z=\begin{bmatrix}
    1&0\\
    0 & -1
  \end{bmatrix}.
\end{equation*}
Let
\begin{equation*}
\sigma_0 = \mathbb{I}_2 =\begin{bmatrix}
    1&0\\
    0 & 1
  \end{bmatrix}
\end{equation*}
be the $2 \times 2$ identity matrix.
For a multi-index $I=(i_1,\ldots,i_m)$, $i_j \in  \{0,x,y,z\}$, let 
\begin{equation} \label{sigmaI}
\sigma_I = \sigma_{i_1}\otimes \cdots \otimes \sigma_{i_m}.
\end{equation}
Any Hamiltonian $H$ may be written as
\begin{equation*}
H=\sum_{I \in \mathcal{I}_H} \alpha_I \sigma_I, 
\end{equation*}
where  $\mathcal{I}_H \subset \{0,x,y,z\}^{\times m}$, and $\alpha_I$ are some real values. 
In an Ising  model  Hamiltonian~\cite{al:ifmnpp},  the multi-index set $\mathcal{I}$ consists of those $I=(i_1,\ldots,i_m)$ with either $1$ or  $2$ of the subindices $i_k = z$, and the remaining $0$, i.e., it has $1$ or $2$ of the  $\sigma_{i_k} = \sigma_z$ in eq.~\eqref{sigmaI} and the rest are $\mathbb{I}_2$.  In a more familiar  notation, an Ising  model Hamiltonian can be written as
\begin{align*} 
H = \sum_{i,j} \alpha_{ij} \sigma^z_i \sigma^z_j + \sum_{k} \beta_{k} \sigma^z_k ,
\end{align*}
where $\alpha_{i j}$ and $\beta_k$ are real, and 
\begin{equation*}
\sigma^z_i=\mathbb{I}_2 \otimes \cdots \otimes \mathbb{I}_2  \otimes \underbrace{\sigma_z}_{\text{at i}}\otimes \mathbb{I}_2  \otimes \cdots  \otimes   \mathbb{I}_2.
\end{equation*}
Let $\theta = (\theta_1,\ldots, \theta_r)$ be a set of real parameters, and let $\ket{\psi(\theta)}$ be a parametrized {\it state} (unit vector) in $\mathcal{H}$. Then a variational quantum algorithm tries to minimize the  mean value of $H$: $\bra{\psi(\theta)} H \ket{\psi(\theta)}$, over the set of values of $\theta$. It is clear that 
\begin{equation*}
\bra{\psi(\theta)} H \ket{\psi(\theta)} \ge \lambda_0,
\end{equation*}
 where $\lambda_0$ is the minimum eigenvalue of $H$. If equality is attained in the above for some set of parameters $\theta^*$, then $\ket{\psi(\theta^*)}$ is an eigenvector of $H$ with the eigenvalue $\lambda_0$.

The quantum approximate optimization algorithm,  QAOA~\cite{fgg:qaoa}, has been shown in~\cite{fgg:qaoa} to approximately find the minimum  (or the maximum as the case may be)  of a class of combinatorial  optimization problems.    As QAOA are a specialized form of the more general VQE, that entails the same for the VQE. Ultimately, it is an attestation of  their convergence to the minimum eigenvalues of Hamiltonians including the type we are concerned with. VQE convergence~\cite{mrb:tvhqca,yrs:ovqapmp} depends on the parametrized state $\ket{\psi(\theta)}$, called the variational ansatz, and the classical optimizer that adjusts the parameters $\theta$ during the search for the minimum eigenvalue (ground state) of the Hamiltonian.  QAOA~\cite{fgg:qaoa,nlc:oqaoasprdcd, zwc:qaoapmin} gains by a state preparation circuit  with an adjustable number ($p$) of  layers of parametrized gates that are based on the Hamiltonian whose ground state is being sought. This structure makes it better adapted to the Hamiltonian and capable of converging to the ground state in the large $p$ limit~\cite{fgg:qaoa, zwc:qaoapmin}.~\footnote{There is no assertion here about the rate of convergence or the approximation error decay with the number of steps of the algorithm, i.e., its complexity.} 
\section{Quadratic Unconstrained Signed  Optimization} \label{sec:quso}
Let us recall the definition of a quadratic unconstrained signed  optimization (QUSO) problem. Let  $Q$ be  a positive symmetric  $m \times m$ matrix.
\begin{defn} [Quadratic unconstrained signed  optimization (QUSO) problem] \label{quadmin}
 Let
\begin{align} \label{quadobj}
h(\omega) = \omega^{\top} Q \omega, ~\text{where } \omega \in \{-1,1\}^{\times m}.
\end{align}
The object of the problem is to find the minimum value $h_{\min} = \underset{\omega \in  \{-1,1\}^{\times m}}{\min} h(\omega)$,
and an element $\omega^*  \in \{-1,1\}^{\times m}$ minimizing $h$, i.e.,  $h(\omega^*) = h_{\min}$.
\end{defn}
Here, $h$ is called the {\it objective function}.
The problem can be written as
\begin{align} \label{quz}
&\text{minimize } \:\: h(\omega) = 2 \sum_{i<j} Q_{i j} \omega_i \omega_j  + \text{tr}(Q), \nonumber \\
&\text{subject to } \: \omega_1, \ldots, \omega_m = \pm 1,
\end{align}
where $Q_{i j}$ is the $i,j$ entry of $Q$. 
This can be transformed to  an Ising Hamiltonian as follows.
Let us consider the complex Hilbert space as in eq.~\eqref{hilbsp},
\begin{equation*} 
\mathcal{H}= {\mathbb{C}^2}^{\otimes m}.
\end{equation*}
We  set aside the constant offset, $\text{tr}(Q)$, in~\eqref{quz},  mapping  the quadratic sum  to an Ising Hamiltonian,
\begin{align} \label{isingham}
H = 2 \sum_{i<j} Q_{i j} \sigma^z_i \sigma^z_j .
\end{align}
Note that $\ket{0}, \ket{1}$ are  the eigenvectors of $\sigma_z$ with the eigenvalues  $1,-1$, respectively. Define the binary variables $y_i= (\omega_i+1)/2$. Then each basis vector of $\mathcal{H}$ of the form 
\begin{equation} \label{kety}
\ket{y}={\bigotimes}_{i=1}^{m} \ket{y_i} = {\bigotimes}_{i=1}^{m}  \ket{\dfrac{\omega_i+1}{2}}
\end{equation}
is an eigenvector of the Ising Hamiltonian $H$. We call such an eigenvector a {\it product-basis eigenvector}. Further, if we denote $\lambda_y$ to be the eigenvalue for the eigenvector $\ket{y}$, then~\footnote{Note that the form $\omega^{\top} Q \omega$ is invariant under the reflection $\omega \mapsto -\omega$, therefore, the map in eq.~\eqref{kety} does not affect the signs  in  eq.~\eqref{lambday}.}   
\begin{equation} \label{lambday}
\lambda_y = \omega^{\top} Q \omega - \text{tr}(Q) = h(\omega) - \text{tr}(Q),
\end{equation}
 where $h$ is the objective function of eq.~\eqref{quadobj}. 
Let us state this  as a theorem,~\footnote{The authors believe that this result is well-known.}

\begin{thm} \label{isingencthm}
Let  $Q$ be an $m\times m$ positive symmetric matrix. Let $h$ be  the objective function in eq.~\eqref{quadobj} and $H$  be the  Ising Hamiltonian  in~\eqref{isingham}, each  constructed from $Q$. Then there is a bijection
\begin{equation*} 
\gamma: (\omega, h(\omega)) \leftrightarrow (y, \lambda_y),
\end{equation*}
where $y$ and $\lambda_y$ are as in eqs.~\eqref{kety} and~\eqref{lambday}, respectively.
\end{thm}
The obvious corollary is
\begin{cor} \label{qusohamcor}
Let $Q$ be an $m\times m$ positive symmetric matrix. A quadratic unconstrained signed  optimization (QUSO)  problem (Definition~\ref{quadmin})  with  $h$ in eq.~\eqref{quadobj} as its objective function   is equivalent to the problem of finding the minimum eigenvalue, and a corresponding  product-basis eigenvector of the Hamiltonian $H$ in~\eqref{isingham}.
\end{cor}


\section{Ising models for Euclidean discrepancy and covariate-balancing} \label{sec:isingeucdisc}

Let $x_1,\dots, x_m \in \mathbb{R}^n$ be an arbitrary set of vectors and let $X$ be the matrix in eq.~\eqref{Xmat} using them as columns. 
Let $Q_X=X^{\top} X$.  We can rewrite the (square of) coloring-discrepancy  in eq.~\eqref{eucdicfunmat} as
\begin{align*} 
h_X=d_X(\omega)^2 = \norm{ \sum_{i=1}^m \omega_i x_i}^2  =  \omega^{\top} Q_X \omega
.
\end{align*}
Then the search for the Euclidean discrepancy given by eq.~\eqref{eucdisc} and an  optimal coloring  satisfying eq.~\eqref{eucdiscopt}, is  a QUSO problem with the objective function  $h_X$. 
Let us convert $h_X$ to an Ising Hamiltonian,
\begin{align} \label{isinghamX}
H_X = 2 \sum_{i<j} {Q_X}_{i j} \sigma^z_i \sigma^z_j .
\end{align}

Covariate-balancing considers the problem of assigning subjects in a randomized trial to two groups so that there is a degree of {\it robustness} against the worst-case error, specified by a parameter, that trades it with balancing the covariates among the groups.  The algorithm in~\cite{hssz:bcregswd} for  covariate balancing in randomized trials, makes use of  the Gram-Schmidt walk algorithm of~\cite{bdgl:gswcbb}.  Let an $n \times m$ matrix $X$ be the  $n$-dimensional covariate (row) vectors   of $m$  subjects in an experimental study.  Assume the vector $\omega \in \{-1,+1\}^{\times m}$ represents the assignment of the  $m$ subjects enumerated by $i = 0,\ldots, m-1$ to one of two possible {\it treatment} groups under the  experiment, grouped by the sign $\omega_i = \pm 1$.  For a random vector $v \in \mathbb{R}^k$, for any $k$,  denote its covariance matrix by $\text{Cov}(v) = \text{E}(v v^\top)$, where $E$ denotes  the expected value  over the distribution from which $v$ is sampled.   The algorithm in~\cite{hssz:bcregswd}  has the flexibility to trade-off  balance, embodied in   $\text{Cov} (X \omega)$,   with robustness, which would  best be attained by making  $\text{Cov}(\omega) = \mathbb{I}_m$.   The $(m+n) \times m$ {\it augmented covariate} matrix $B$ in~\cite{hssz:bcregswd} trades off these through a parameter $\phi$,
\begin{equation*} 
B = \begin{bmatrix}
\sqrt{\phi} \mathbb{I}_m \\
   \xi^{-1} \sqrt{1-\phi} X
  \end{bmatrix},
\end{equation*}
where $\mathbb{I}_m$ is an $m\times m$ identity matrix.  The parameter  $\phi \in [0,1]$ determines the degree to which the vector $\omega$ approximates either the result of {\it raw covariates} given by $X$ or  that from the  pairwise independent assignments coming from $\mathbb{I}_m$. The parameter $\xi = \max_{i \in [m]} \norm{x_i}$  ensures that  the two matrices have  equal influence in contributing to $\omega$. For the rest of this section,  fix the parameter $\phi$. We will call the colorings $\omega$ that minimize $\norm{B \omega}$ as optimal covariate-balancing colorings for the matrix $B$. 

Write $B$ as column vectors $b_1,\ldots, b_m \in  \mathbb{R}^{m+n}$,
\begin{equation*} 
B=\begin{bmatrix}
    \vline & \cdots & \vline  \\
     b_1& \cdots  & b_m\\
     \vline & \cdots  &  \vline 
  \end{bmatrix},
\end{equation*}
where 
\begin{equation*} 
b_i= 
  \begin{bmatrix}
	\sqrt{\phi} e_i \\
	\xi^{-1} \sqrt{1-\phi} x_i\\
  \end{bmatrix},  
\end{equation*}  
and $e_i$ is the $i^{\text{th}}$ standard coordinate vector in $\mathbb{R}^m$.
Define the {\it assignment-imbalance},
 \begin{align*} 
i_X &: \{-1,1\}^{\times m} \rightarrow \mathbb{R} \nonumber \\
i_X(\omega) &=   \norm{ \sum_{i=1}^m \omega_i b_i}
\end{align*}
%
Taking the minimum over $\omega$ of the assignment-imbalance,  we can define the  {\it augmented-imbalance}~\footnote{The authors are unaware of the proper terminology here.} as
 \begin{equation} \label{imbal}
\text{imb}(X) = \min_{\{\omega\}}i_X(\omega).
\end{equation}
An {\it optimal assignment}  
 $\omega^*$ as one that achieves the augmented-imbalance,
 \begin{equation} \label{covimbalopt}
i_X(\omega^*) = \text{imb}(X).
\end{equation}
And as in the case of the Euclidean discrepancy, define the matrix
\begin{equation*}
Q_B=B^{\top} B.
\end{equation*}  
We can rewrite the  square of the assignment-imbalance  as
\begin{equation*} 
h_B(\omega) = i_X(\omega)^2 =  \omega^{\top} Q_B \omega.
\end{equation*}

The problem of finding an optimal assignment and the augmented-imbalance is then a QUSO problem with the objective function $h_B$.
Converting $h_B$ to an Ising Hamiltonian, we have
\begin{align} \label{isinghamB}
H_B = 2 \sum_{i<j} {Q_B}_{i j} \sigma^z_i \sigma^z_j .
\end{align}

Then, by Corollary~\ref{qusohamcor}, we have
\begin{thm} \label{euccovising}
Given a set of vectors $x_1,\dots, x_m \in \mathbb{R}^n$,
 \begin{enumerate} [label=(\alph{*})] 
 \item \label{ecia}   finding their Euclidean discrepancy given by eq.~\eqref{eucdisc} and an optimal coloring given by eq.~\eqref{eucdiscopt} is equivalent to finding the minimum eigenvalue, and a corresponding   product-basis  eigenvector of the Ising Hamiltonian $H_X$ in eq.~\eqref{isinghamX}.  
  \item \label{ecib}  finding their augmented-imbalance given by eq.~\eqref{imbal}, and an optimal assignment given by eq.~\eqref{covimbalopt} is equivalent to finding the minimum eigenvalue, and a corresponding  product-basis eigenvector of the Ising Hamiltonian $H_B$ in eq.~\eqref{isinghamB}.
  \end{enumerate}
\end{thm}

The constructions and results  in~\cite{hssz:bcregswd} provide  extensive analysis,  bounds and criteria quantifying performance,  implications and complexity of the algorithm. From a  much simplified and coarse perspective, the algorithm  samples from $\omega$ such that  the largest eigenvalue of  $\text{Cov} (B \omega)$ is close to the minimum that it can be over all $\omega  \in \{-1,1\}^{\times m}$.

Let us see how this behaves  in the variational quantum algorithms case.  A variational quantum optimizer samples the vector  $\omega$ (equivalently $\ket{y}$ in eq.~\eqref{kety}) from the values that make $\text{E}(\norm{B \omega}^2)$ close to the minimum. A useful observation is that  $\norm{B \omega}^2 = \text{tr} (B \omega  \omega^{\top} B^{\top})$, which implies that $\text{E}(\norm{B \omega}^2) = \text{tr } \text{Cov}(B\omega) = \sum_{k=1}^n \lambda_k$, where $\lambda_k$ are the eigenvalues of $\text{Cov}(B \omega)$. This  means that the largest eigenvalue of  $\text{Cov} (B \omega)$ will be close to the minimum it can be  over all  $\omega  \in \{-1,1\}^{\times m}$.

 When $\phi=0$, $Q_B=X^{\top} X$, which is  the same as in the previous section.  When  $\phi=1$,  $Q_B=\mathbb{I}_m$. All of the eigenvalues of $Q_B$ are $1$. Then a VQE capable of sampling from the entire space,  would sample $\omega$  uniformly from all its possible values, resulting in  $\text{Cov}(B\omega) = \text{Cov}(\omega)= \mathbb{I}_m$.

Sometimes it might be  desirable to have the same number of $+1$'s as $-1$'s in the optimal assignment, i.e., that the number of subjects are split equally into the treatment groups. This is called the {\it group-balanced randomization design} in~\cite{hssz:bcregswd}.  For that,  the particle-number preserving ansatz from~\cite{gzb:espspcvqea}, with $m/2$    particles,  may be used in the variational algorithm. We leave the details of the proper ansatz choice from the cited work to the interested reader. In the next section, we test the VQA on an IBM quantum device simulator to see if they work as expected.

\section{Running variational quantum algorithms  on quantum simulators} \label{sec:quantsim}

Here we demonstrate, using a simple example, the variational quantum algorithm for covariate balancing. We run the algorithm on IBM quantum (qasm) simulator~\cite{ibm:qexp} with the code written in the IBM Qiskit~\cite{qiskiturl} framework,~\footnote{The code is available upon request.} and compare the result with the algorithm in~\cite{hssz:bcregswd}. The code for the algorithm in~\cite{hssz:bcregswd} is available as a repository~\cite{hs:gswd}. 
Publicly accessible IBM quantum simulators have the VQE and QAOA routines available, and we use them, indicating the parameter settings, but without describing what they mean. The reader is referred to the online resources within IBM Qiskit~\cite{qiskiturl}, or to the several papers we have cited on the topics~\cite{pms:vqepqp,mrb:tvhqca,fgg:qaoa,nlc:oqaoasprdcd, pbb:qaoalrimtqs,zwc:qaoapmin}. 

The experiment we conduct is limited by the resources available. The dimensions of the multi-qubit Hilbert space grows exponentially with $m$, the number of covariate vectors, i.e., the number of subjects in the trial. The purpose of this experiment is to demonstrate the variational quantum algorithms approach. We generate $m=12$  covariate vectors, $\{x_i \in \mathbb{R}^2\}$, drawn evenly from $2$ Gaussian distributions with means $[-3,3]$ and $[3,3]$, and variances   $1$ in each coordinate direction. These are shown in figure~\ref{fig1}, and given as the matrix $X$ in Appendix~\ref{appdx:data}, eq.~\eqref{Xcovmat}. The parameter  balance-robustness parameter $\phi = 0.5$ for all the assignments that follow, except, naturally,  for the uniformly random assignment.
\begin{figure}[H]
\includegraphics[scale=0.5]{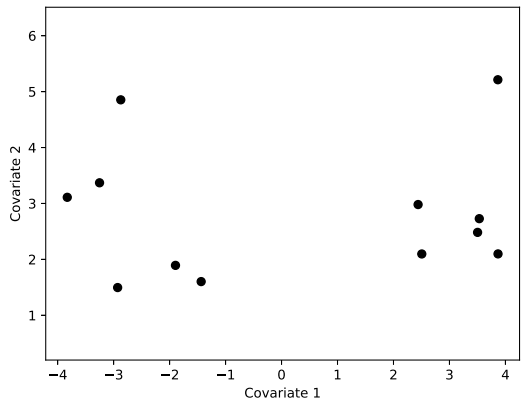}
  \caption{
    Covariate vectors from $2$ Gaussian distributions.
  }
  \label{fig1}
\end{figure}
 Let us assign these vectors to two groups, i.e., sample the assignment vector $\omega \in \{\pm 1\}^{\times m}$ using some randomized method. 

\subsection*{Uniformly random}

\begin{figure}[H]
\includegraphics[scale=0.5]{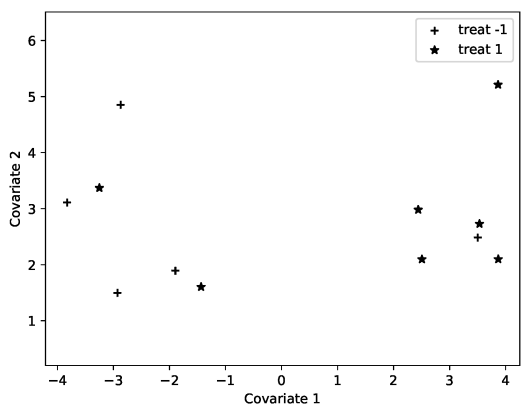}
  \caption{
    Uniformly random assignment 
  }
  \label{fig2}
\end{figure}

The entries $\omega_i = \pm 1$ are chosen independently and uniformly randomly. The sampled $\omega$ is in eq.~\eqref{omrandom}. Figure~\ref{fig2} shows the assignment of each vector $x_i$ to the corresponding treatment group $\omega_i$. We observe that the each group is ``clumped" in one of the two distributions. 

\subsection*{GSW}

\begin{figure}[H]
\includegraphics[scale=0.5]{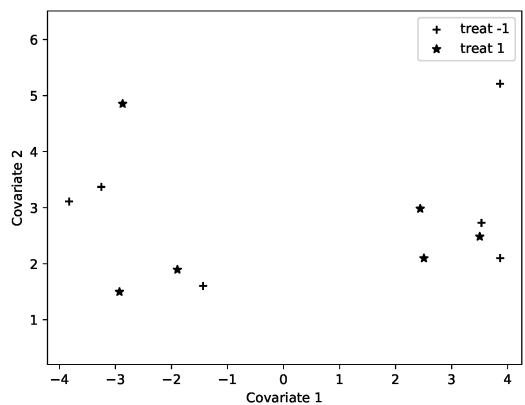}
  \caption{
  GSW assignment
  }
  \label{fig3}
\end{figure}

Next we sample the assignment based on the GSW algorithm described in\cite{hssz:bcregswd} implemented in the code  repository~\cite{hs:gswd}. We obtain a sample $\omega$ given  in eq.~\eqref{omgsw}, after a few runs that attains the best observed  assignment-imbalance  value  $i_X(\omega) = 2.4720$. 
The assignment to treatment groups is shown in figure~\ref{fig3}. 

\subsection*{VQE}
\begin{figure}[H]
\includegraphics[scale=0.5]{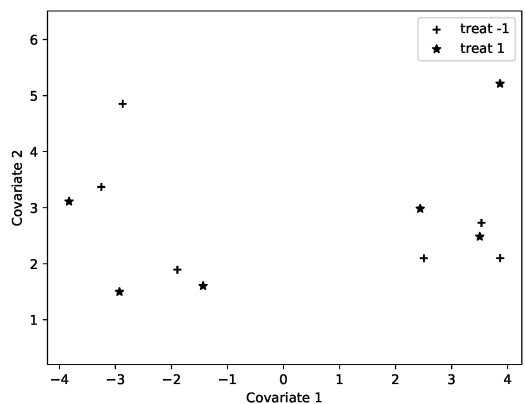}
  \caption{
VQE assignment
  }
  \label{fig4}
\end{figure}

We test the VQE algorithm of IBM Qiskit~\cite{qiskiturl} with the setup of Section~\ref{sec:isingeucdisc}.~\footnote{The simulation runs $65536$ shots on the qasm-simulator backend.  It uses  the two-local variational ansatz,  with rotation-blocks:  [``rz",``ry"],  entanglement-blocks: ``cx", reps: $3$, entanglement: ``circular". The rest of the parameters are the default ones. The optimizer used is the Constrained Optimization by Linear Approximation optimizer (COBYLA). }
From VQE we get a sample $\omega$ given in eq.~\eqref{omvqe}, 
with the assignment-imbalance  value  $i_X(\omega) = 2.4497$. Figure~\ref{fig4} shows the sampled assignment.  

\subsection*{QAOA}
\begin{figure}[H]
\includegraphics[scale=0.5]{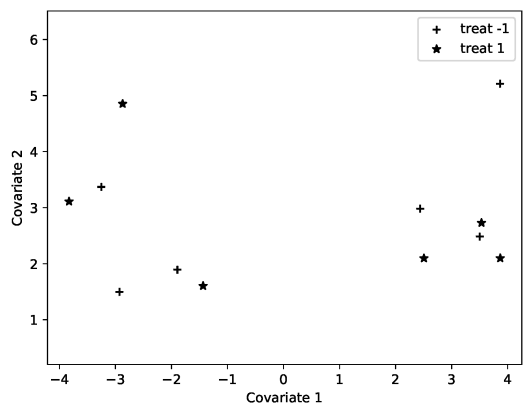}
  \caption{
  QAOA assignment
  }
  \label{fig5}
\end{figure}

We test the QAOA algorithm of IBM Qiskit~\cite{qiskiturl},  for the same setup as for VQE.~\footnote{This simulation runs $65536$ shots on the qasm-simulator backend.  It uses $p=8$, with a uniform superposition initial state. The optimizer used is the Constrained Optimization by Linear Approximation optimizer (COBYLA).}
This yields a sample $\omega$ given in eq.~\eqref{omqaoa}, 
with the assignment-imbalance  value  $i_X(\omega) =  2.4516$. The assignment is displayed in figure~\ref{fig5}.

\subsection*{Optimal by exhaustive search}
\begin{figure}[H]
\includegraphics[scale=0.5]{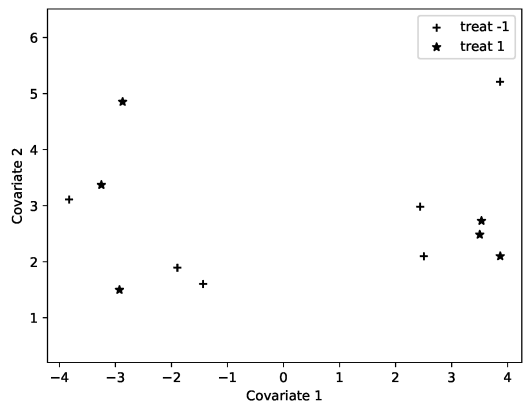}
  \caption{
Optimal assignment
  }
  \label{fig6}
\end{figure}

As a comparison, we also find, through exhaustive search, the imbalance $\text{imb}(X)$ (minimum assignment-imbalance  value), and an optimal assignment. The optimal  $\omega$  we find is given in eq.~\eqref{omes}, 
and the imbalance is $\text{imb}(X) = i_X(\omega^*) =   2.4496$. Figure~\ref{fig6} shows the optimal assignment. 

\section{Conclusion} \label{sec:conc}
 Posing the classical problems of Euclidean discrepancy and covariate balancing as QUSO problems allows us to convert them to  Ising models.  This turns the search for optimal colorings and assignments of subjects in randomized trials to the search for the lowest eigenvalues of Ising Hamiltonians. These are precisely the problems for which the VQA, in particular  the VQE and the QAOA, have been developed and furnish powerful approaches to optimization in the NISQ regime of quantum computing. The experiment in the previous section demonstrates the efficacy of using the VQA in this search, whose performances  compare favorably, though for a modest example,  to the GSW algorithm. We realize the need for extensive tests, especially  on actual quantum devices, to draw meaningful comparisons. Within covariate-balancing, different parameters values could be tested, and so could  constrained optimizations  such as the  group-balanced randomization design, for which we cited a possible VQE ansatz from the extant literature.   Future work may seek to  expand the VQA  to  other aspects of algorithmic discrepancy theory. Another interesting, and indeed potentially unexplored, direction is the investigation of  efficient quantum algorithms for discrepancy theory, with the recent discoveries in classical algorithms for it serving as a motivation and as possible guides.  

\section*{Acknowledgments} \label{sec:ack} The authors would like to thank Daniel Spielman for encouraging them to think about quantum counterparts of algorithmic  discrepancy theory.

\appendix
\section{Numerical simulation data} \label{appdx:data}


Matrix of covariates (transpose):
\begin{align}
 \label{Xcovmat}
X^{\top}=\begin{bmatrix}
  3.8673 &  2.0983\\
  2.5055 &  2.0971\\
  3.8644 &  5.2119\\
  3.5328 &  2.7283\\
  3.5023 &  2.4830\\
  2.4395 &  2.9807\\
 -2.8719 &  4.8528\\
 -3.8278 &  3.1101\\
 -3.2512 &  3.3697\\
 -2.9279 &  1.4966\\
 -1.4358 &  1.6033\\
 -1.8945 &  1.8933
\end{bmatrix}.
\end{align}

Uniformly randomly sampled $\omega$:
\begin{equation} \label{omrandom}
\omega = \begin{bmatrix}
  1  ,  1  ,  1  ,  1  , -1  ,  1  , -1  , -1  ,  1  , -1  ,  1  , -1 
\end{bmatrix}.
\end{equation}

GSW sampled $\omega$:
\begin{equation} \label{omgsw}
\omega = \begin{bmatrix}
1  ,  1  , -1  ,  1  , -1  , -1  ,  1  ,  1  , -1  , -1  ,  1  , -1
\end{bmatrix}.
\end{equation}

VQE sampled $\omega$:
\begin{equation} \label{omvqe}
\omega = \begin{bmatrix}
-1  , -1  ,  1  , -1  ,  1  ,  1  , -1  ,  1  , -1  ,  1  ,  1  , -1
\end{bmatrix}.
\end{equation}

QAOA sampled $\omega$:
\begin{equation} \label{omqaoa}
\omega = \begin{bmatrix}
-1  , -1  ,  1  , -1  ,  1  ,  1  , -1  ,  1  , -1  ,  1  ,  1  , -1
\end{bmatrix}.
\end{equation}

Optimal $\omega^*$ by exhaustive search:
\begin{equation} \label{omes}
\omega^* = \begin{bmatrix}
1  , -1  , -1  ,  1  ,  1  , -1  ,  1  , -1  ,  1  ,  1  , -1  , -1 
\end{bmatrix}.
\end{equation}

\end{document}